\documentclass[pre,aps,amsmath,eqsecnum,unsortedaddress,nofootinbib,twocolumns]{revtex4-1}
\usepackage{amssymb}
\usepackage{amsmath} 
\usepackage{graphicx}
\usepackage{amsbsy}
\usepackage{color}
\usepackage{bm}
\usepackage{soul}

\begin{document}
\title{Chiral Self-Assembly of Helical Particles}

\author{Hima Bindu Kolli} 
\affiliation{Dipartimento di Scienze Molecolari e Nanosistemi, 
Universit\`{a} Ca' Foscari di Venezia,
Campus Scientifico, Edificio Alfa,
via Torino 155, 30170 Venezia Mestre, Italy}
\affiliation{present address: Department of Chemistry, University of Oslo,
Postboks 1033 Blindern 0315 Oslo, Norway}

\author{Giorgio Cinacchi} 
\email{giorgio.cinacchi@uam.es}
\affiliation{
Departamento de F\'{i}sica Te\'{o}rica de la Materia Condensada, 
Instituto de F\'{i}sica de la Materia Condensada (IFIMAC) and
Instituto de Ciencias de Materiales "Nicol\'{a}s Cabreras", 
Universidad Aut\'{o}noma de Madrid,
Campus de Cantoblanco, 28049 Madrid, Spain
}
\author{Alberta Ferrarini} 
\email{alberta.ferrarini@unipd.it}
\affiliation{Dipartimento di Scienze Chimiche, 
Universit\`{a} di Padova, via F. Marzolo 1, 35131 Padova, Italy}

\author{Achille Giacometti} 
\email{achille.giacometti@unive.it}
\affiliation{Dipartimento di Scienze Molecolari e Nanosistemi, 
Universit\`{a} Ca' Foscari di Venezia,
Campus Scientifico, Edificio Alfa,
via Torino 155, 30170 Venezia Mestre, Italy}

\date{\today}

\begin{abstract}
The shape of the building blocks plays a crucial role in 
directing self-assembly towards desired architectures. 
Out of the many different shapes, helix has a unique position. 
Helical structures are ubiquitous in nature and 
a helical shape is exhibited by  the most important biopolymers like 
polynucleotides, polypeptides and polysaccharides
as well as by cellular organelles like flagella. 
Helical particles can self-assemble into chiral superstructures, 
which may have a variety of applications, e.g. as photonic (meta)materials.  
However,a clear and definite understanding of these structures has not been entirely achieved yet.
We have recently undertaken an extensive investigation on the phase behaviour of hard helical particles, 
using numerical simulations  and  classical density functional theory. 
Here we present a detailed study of the phase diagram of hard helices as a function of their morphology. 
This includes a variety of liquid-crystal phases, 
with different degrees of orientational and positional ordering. 
We show how, by tuning the helix parameters, 
it is possible to control the organization of the system. 
Starting from slender helices, whose phase behaviour is similar to that of rodlike particles,
the increase in  curliness leads to the onset of azimuthal correlations between the 
particles and the formation of phases specific to helices.
These phases feature a new kind of screw order, 
of which there is experimental evidence in 
colloidal suspensions of 
helical flagella.
\end{abstract}
\maketitle
\section{Introduction}
Self-assembly refers to a process where 
multiple components spontaneously organize to form a large entity \cite{Whitesides02}. 
It is viewed as one of the most promising
route to a ``bottom-up'' engineering of new nanomaterials. 
Recent developments, both numerical \cite{Damasceno12} and experimental \cite{Sacanna13,Sacanna13bis}, include the possibility of obtaining a variety of complex structures using  
building blocks of essentially any possible shape.  
While the idea of exploiting the shape of the building blocks as a way to direct the self-assembly process towards the desired architectures has been vigorously
pursued in the last few years \cite{Glotzer07,Cademartiri12}, 
the idea of exploiting also their \textit{chirality} has so far received less attention.

Yet the control of the chiral organization of materials is of major interest for application purposes.
The promising field of  chiral photonics aims at exploiting 
the unique optical properties of chiral (meta)materials, 
like giant gyrotropy, unusually large optical activity and circular dichroism, 
with associated negative refractive index,  or strong nonlinear chiroptical effects  
(for recent reviews see e.g. Refs. \cite{Soukoulis2011, Liz2011, Wegener2015}).
The early observation of an optical effect in a twisted structure was the detection of  rotation of the plane of polarization of microwaves traversing a jute bundle by Bose in 1898 \cite{Bose}. 
In the last few years there has been a renewed interest in this topic and increasing efforts have been devoted to the design, synthesis and characterization of  nano-engineered materials, made of chiral building blocks that may be viewed as "artificial chiral molecules". 
In this context, a special place is deserved to 
helical particles: 
as early as 1920  Lindman showed the polarization rotation of the microwave radiation transmitted through a system of randomly oriented metallic helices  \cite{Lindman}, in analogy with the behavior exhibited by chiral molecules. This can be taken as the origin of helix-based chiro-optical metamaterials: special chiro-optical properties orders of magnitude stronger than those of molecular systems  were recently discovered in ordered arrays of helical particles  \cite{Gansel09}.  
Thanks to the progress in nanofabrication, 
also arrays of plasmonic helices active in the visible wavelength region could be obtained \cite{Larsen,Esposito15}.
Helically nanostructured materials are attracting considerable attention also because of unique electrical and mechanical properties (see  e.g. Refs. \cite{Yang,Fisher,Liu14} for recent reviews). 
The control of the morphology of individual helices  and their 3D organization, which determine the interaction with electromagnetic fields \cite{Gansel09} and may affect the   electrical and  mechanical response of the (meta)materials, is a major step towards the optimization of structures for the different applications. 

The importance of the helix in nature need not be stressed:  a helical shape is exhibited by the most important biopolymers like polynucleotides, polypeptides and polysaccharides, as well as by biological particles like filamentous viruses and cellular organelles like flagella. 
These systems generally exhibit one or more liquid-crystal phases above a given density. The most common is the cholesteric  (N$^\ast$), which is a twisted nematic phase. 
It is nematic since  the centers of mass of helices are randomly distributed in space, but the helical axes are locally preferentially aligned  to one another;  
it is twisted because the alignment direction (the director $\widehat{\mathbf{n}}$) rotates in helical way around a perpendicular axis. Typical periodicities (the cholesteric pitch $\cal P$) may range from micro- to milli-meters, depending on the specific system, but in all cases are orders of magnitude longer than the pitches ($p$) of the constituent helices. 
For some systems, like DNA \cite{Livolant}, in a restricted density range just below the N$^\ast$ phase, 
Blue Phases \cite{Wright89} have also been found. 
These can be described as fluid lattices of defects, with cubic symmetry and   
lattice periods of the order of the wavelength of visible light. 
They are locally chiral, since directors are locally arranged in double-twist cylinders.
On the upper boundary of the  N$^\ast$ phase, smectic phases (Sm) can be found, 
which, beside orientational order, exhibit partial positional order: 
the particles form layers,  but within a layer  positional order is short-range or absent altogether.
In 
polydisperse systems the formation of Sm phases is inhibited and columnar phases may form, 
where arrays of close packed columns are formed by axially stacked particles, 
with continuous translational symmetry along  the axial dimension.

All these phases are not specific to helices and can be found in other chiral or non-chiral systems, and
it was tacitly assumed that nothing special was associated with the helical shape of particles. 
However, this belief was challenged by a recent study of dense colloidal suspensions of helical flagella,  
which are strongly curled helical filaments, with a well defined pitch  $p \sim 1-2 \mu$m \cite{Barry06}.  
To explain the results of polarizing and differential interference contrast microscopy experiments,
a novel chiral nematic organization was proposed, 
where helical filaments would be preferentially aligned along a common direction, 
intercalated and in-phase with each other without positional order.  
This picture is compatible with the observation of a striped birefringent pattern with 
periodicity  equal to the helix pitch $p$, 
since in such a structure the local tangents to helices undergo a conical rotation with 
period $p$ around the alignment direction. 
Then, a number of questions arise: why there are no evidences of this phase in other helical systems? 
Are there special requirements for helices to form this new phase or is it only a matter of detection? 
In fact, whereas a periodicity in the micrometer range is immediately evidenced by polarization microscopy, 
analogously direct methods are not currently available  in the nanometer range, 
as are the helical pitches of most (bio)polymers.
A more general question is whether self-assembly of helical particles may lead to 
other special phases, not yet discovered or overlooked.

Clearly, both fundamental and application questions call for 
a comprehensive  insight into
the self-assembly of helical particles.  
This issue remained scarcely explored,
which is surprising in view of the paradigmatic role of the helical shape.
To fill this gap, we recently undertook a study
of self-assembling hard rigid helices, using Monte Carlo simulations and classical density functional theory (DFT) 
at the second-virial (Onsager) and third-virial level \cite{Frezza13,Frezza14,Kolli14a,Kolli14b}.
The choice of hard particles, thus interacting through a purely repulsive short-range potential, 
was dictated by the intent of exploring the effects determined solely by the helical shape.  
We found a rich polymorphism, with the presence of 
chiral liquid-crystal phases not previously observed 
for other chiral systems.
A remarkable result \cite{Kolli14a} was the detection of a chiral  phase,  
called screw-nematic \cite{Manna07},
which has the same features of the phase detected  in  
dense suspensions of helical flagella\cite{Barry06}.     
On increasing densities more, 
we found smectic phases, generally chiral, 
which also are specific to helical particles. 
These are followed, at very high density, by chiral compact phases.  

\begin{figure}[htbp]
\begin{center}
\includegraphics[width=1.5in]{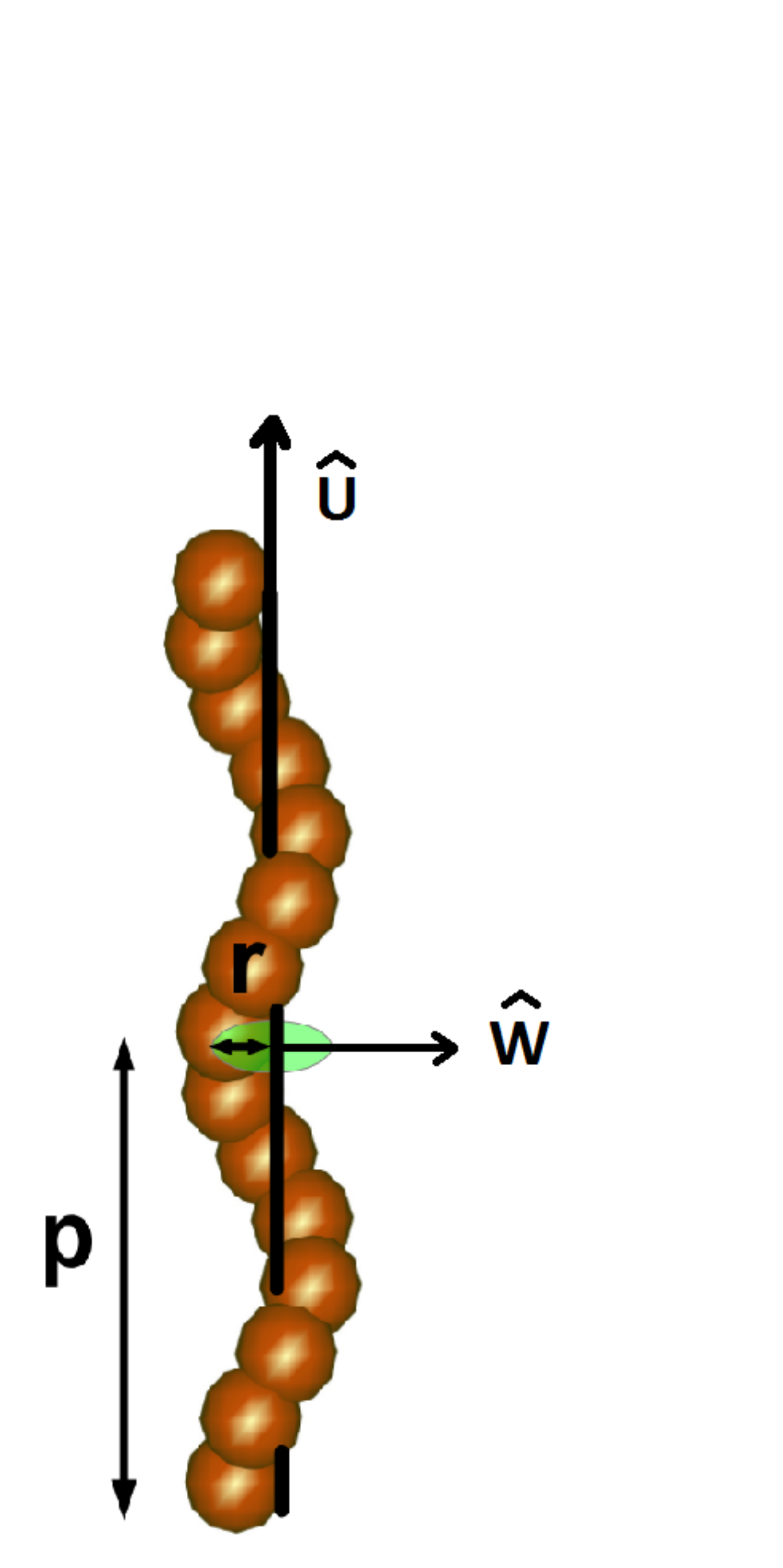}
\hfill
\hspace{0.2in}
\includegraphics[width=3.0in]{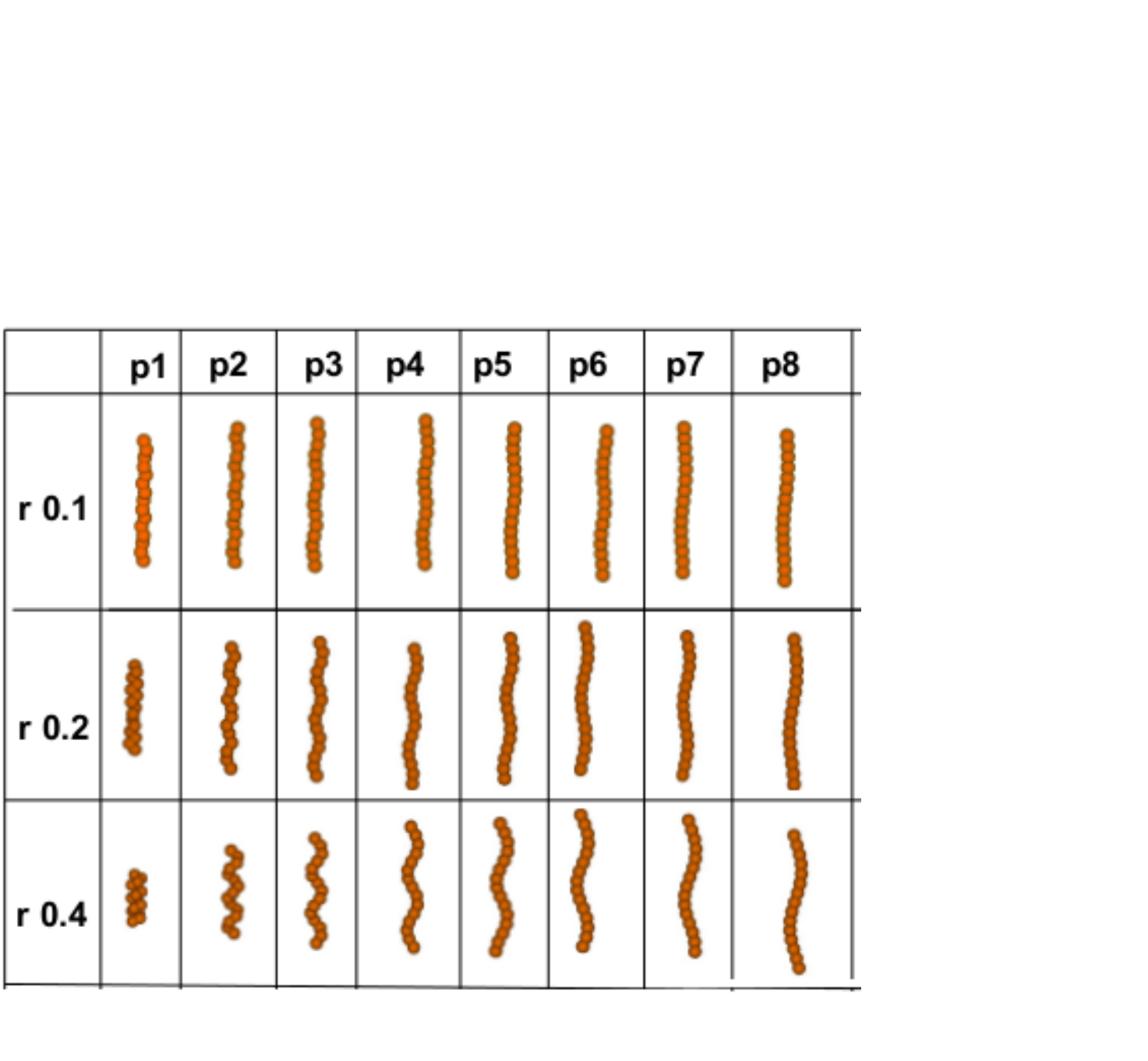}
\end{center}  
\caption{(Left) Model helical particle with  prescribed helix radius $r$ and pitch $p$. 
Unit vectors $\widehat{\mathbf{u}}$
and $\widehat{\mathbf{w}}$ are parallel to the helical axis and to the two-fold symmetry axis, respectively.
(Right) Morphology of the helix models used in this work:  
$r$ is the helix radius and $p$ is the pitch. Different combinations of $(r,p)$ then give rise to different shapes and aspect ratios. }
\label{fig:fig1}
\end{figure}

In our studies, we used the helix model shown in Fig. \ref{fig:fig1}(Left):
each particle is  composed of 15 partially fused hard spheres of diameter $D$, 
rigidly arranged in a helical fashion with a fixed contour length 
$L=10D$.
Hence  different helix morphologies can be achieved upon independent change of 
the radius $r$ and the pitch $p$. 
Here the bead diameter is a scaling parameter and the phase behaviour  
is determined by the values of $r$ and $p$ relative to $D$: 
for $p<2D$ helices cannot intrude into each other grooves and, irrespective of $r$, resemble threaded rods.   
The mutual helical character of particles clearly emerges only for $(r,p)$ bigger than the sphere diameter. 
Hereafter, the bead diameter $D$ is taken as the unit of length.
In the previous Monte Carlo simulations,  we focused on 
three  different morphologies ($r=0.2$, $p=8$; $r=0.2$, $p=4$; $r=0.4$, $p=4$), 
all with clear helical character,
and we identified  the possible phases formed by helical particles \cite{Kolli14a,Kolli14b}. 
Here, with a considerable computational effort, 
we have extended our investigation to a wider range of $(r,p)$ values, 
to get a more detailed  insight into the phase behaviour
as a function of the helical morphology. The differences 
in terms of combinations of radii and pitches is rather remarkable, as illustrated in Fig.\ref{fig:fig1} (Right). 
The two extreme cases are represented by $r=0.1$, where helices are rather tube-like for any pitch value, and by $r=0.4$, where different pitches lead to significantly different morphologies and aspect ratios.
The low chirality, $r=0.1$, systems will allow us to investigate a region that remained hitherto unexplored, which however deserves to be considered, as representative of the behavior of systems like (bio)polymers.
The case $r=0.2$ present intermediate phenomenologies.  
The analysis of the phase diagram will show how the organization of the systems can be controlled by tuning the helical character 
of the building blocks. 
 
The remaining of the paper is organized as follows. In Sec. \ref{sec:methods} we briefly recall the 
methods used in our study, whereas in Sec. \ref{sec:order} we review the
definitions of the order parameters and correlation functions necessary to identify the various phases. 
Sec. \ref{sec:phase} contains the results for the phase diagrams and 
Sec. \ref{sec:conclusions} the conclusions and possible perspectives.    

\section{Methods}
\label{sec:methods}
Isobaric Monte Carlo simulations have been exploited to compute the equation of state. 
The interested reader is referred to previous works for details \cite{Frezza13, Kolli14a,Kolli14b}. 
Here we shall recall only the few points necessary for the sake of completeness.

A variable number $N$, ranging between 900 and 2000, of hard helices, were used 
depending on the helix morphologies and the required densities. 
In all cases, the initial conditions were taken as an ordered compact configuration generated 
using the isopointal search method described in Ref. \cite{Kolli14b}, 
but the stability of the obtained results with respect to a change in the initial conditions was 
explicitly tested for.
A typical Monte Carlo cycle consisted of $N/2$ translation moves, $N/2$ rotation moves, as well as one volume move, with a typical equilibration time of the order of $3-4 \times 10^6$ Monte Carlo steps,
as well as productions runs of  additional $1-2 \times 10^6$ Monte Carlo steps. 

In all simulations periodic boundary conditions (PBC) were used,
with a floppy (i.e. shape adaptable) triclinic computational box. Such conditions are
fully compatible with the existence of helical order with small periodicity, comparable with the particle length. 
However they do not  allow  the development of a cholesteric organization 
when this is characterized by a periodicity that is 
orders of magnitude longer than the scale of particle chirality \cite{notarella}.  
For this reason using PBC in our simulations we could only find N phases, 
with uniform rather than twisted $\widehat{\mathbf{n}}$ director (i.e. $\cal{P} \rightarrow \infty$). 
To ascertain the possible existence of a director twist, we subsequently
evaluated the pitch $\cal P$ for all state points in the putative N$^\ast$ phases, using DFT 
within the Onsager second-virial \cite{Onsager} framework with Parsons-Lee correction \cite{parsons-lee}. 
The method presented earlier \cite{Tombolato05,Frezza14} was adopted and 
calculations were performed according to the following procedure: 
(i) The order parameters at a given density were determined by minimization of the free energy of the untwisted N phase; 
(ii) These order parameters were used to calculate the chiral strength $k_2$ and the twist elastic constant $K_{22}$; 
(iii) The cholesteric pitch was determined as ${\cal P}=-k_2/K_{22}$. 
The elastic constant is a positive quantity, whereas the chiral strength, and thus the cholesteric pitch  ${\cal P}$, 
are signed quantities (with opposite signs for enantiomers). 
Positive and negative ${\cal P}$ correspond to a right-handed and a left-handed cholesteric phase, respectively. 
 
\section{Order parameters and correlation functions}
\label{sec:order}
We used a combination of order parameters and correlation functions to 
fully characterize the various phases encountered in this study. 
Here we will briefly recall the main definitions, referring to Refs. \cite{Kolli14a,Kolli14b} for details.

The orientation of a  helix in space is univocally identified by the unit vectors $\widehat{\mathbf{u}}$ and $\widehat{\mathbf{w}}$, 
parallel to the main axis and to the $C_2$ symmetry axis, respectively 
(see Fig. \ref{fig:fig1}).
Nematic order is identified by the order parameter 
\begin{eqnarray}
\label{eq1}
\left \langle P_2 \right \rangle &=& 
\left \langle \frac{3}{2} (\widehat{\mathbf{u}} \cdot \widehat{\mathbf{n}})^2 - \frac{1}{2} \right \rangle,
\end{eqnarray}
which describes the average (non polar) alignment of the $\widehat{\mathbf{u}}$ axes of helices to the $\widehat{\mathbf{n}}$ director and vanishes in the isotropic state. Here and henceforth  the angular brackets denote ensemble averages.    
The $\widehat{\mathbf{n}}$ director and the $\left \langle P_2 \right \rangle$ order parameter are computed according to the 
Veilliard-Baron procedure \cite{Veilliard}.
 
Transversal polar order is quantified by the order parameter
\begin{eqnarray}
\label{eq2}
\left \langle P_{1,c} \right \rangle &=& \left \langle \widehat{\mathbf{w}} \cdot \widehat{\mathbf{c}} \right \rangle,
\end{eqnarray}
which measures the average alignment  of the $\widehat{\mathbf{w}}$ axes of helices along a common direction ($\widehat{\mathbf{c}}$). In the case of screw order the  $\widehat{\mathbf{c}}$ axis rotates in helical way on moving along the $\widehat{\mathbf{n}}$ director, with pitch $\cal P$ equal to the particle pitch $p$ \cite{Kolli14a,Kolli14b}.

\begin{figure}[htbp]
\begin{center}
\includegraphics[width=2.5in]{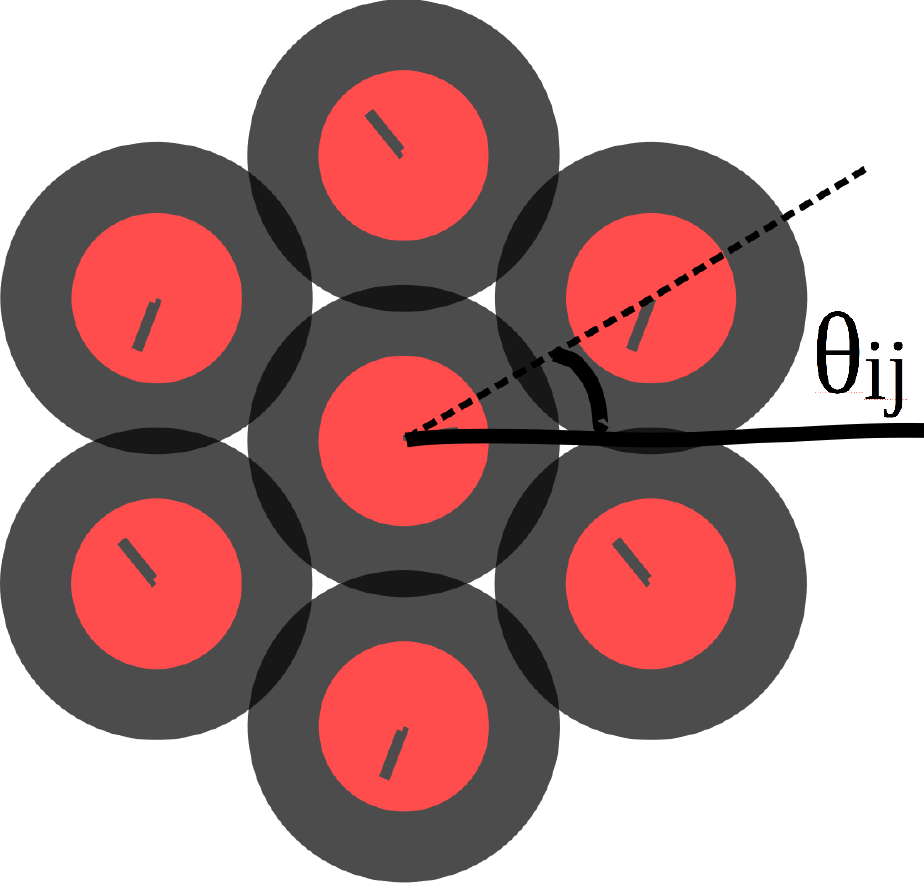}
\end{center}  
\caption{Top view representation of short-range hexagonal ordering in helices. The dashed line joining the $(i,j)$ nearest-neighbour helices
makes an angle $\theta_{ij}$ with a reference axis (solid line). Helices are projected onto a plane perpendicular to the nematic director  $\widehat{\mathbf{n}}$.
Red circles denote the radius of the helix, while the thick short line inside each circle indicate the orientation of the helix azimuthal direction $\widehat{\mathbf{w}}$
associated with the $C_2$ symmetry  axis.}
\label{fig:fig2}
\end{figure}

Smectic order, with the formation of layers perpendicular to the director $\widehat{\mathbf{n}}$, is monitored by using the  order parameter
\begin{eqnarray}
\label{eq3}
\left \langle 
\tau_1 
\right \rangle 
&=& \left | \left \langle \mathsf{e}^{i 2\pi \frac{\mathbf{R}\cdot\widehat{\mathbf{n}}}{d}}  
\right \rangle \right |,
\end{eqnarray}
where $\mathbf{R}$ is the position of particle centre of mass and $d$ the layer spacing. 
The most common smectic phases are those denoted as  A (Sm$_{A}$) and B (Sm$_{B}$), 
which are both uniaxial and differ in the in-plane positional order, 
which is absent in the former, whereas in the latter there is short-range in-plane hexagonal order.   
This can be probed by using the hexagonal order parameter
\begin{eqnarray}
\label{eq4}
\left \langle \psi_{6} \right \rangle &=& \left \langle \frac{1}{N} \sum_{i=1}^N \left \vert \frac{1}{n(i)} \sum_{j=1}^{n(i)} e^{6 \mathrm{i} \theta_{ij}}
\right \vert \right \rangle.
\end{eqnarray}
where $n(i)$ is the number of in-layer nearest-neighbors of the $i$-th helix, and $\theta_{ij}$ is the angle between the $i,j$ intermolecular vector and a reference axis in the plane perpendicular to $\widehat{\mathbf{n}}$ , 
as illustrated in Fig. \ref{fig:fig2}. 
Then $\langle \psi_{6} \rangle$ takes on a non-zero value in the presence of a short-range hexagonal order, and vanishes otherwise. 

In Ref. \cite{Kolli14b} new smectic phases were identified for helical-particle systems. 
The screw-smectic A (Sm$_{A,s}$) is a uniaxial smectic phase with screw order within planes, 
quantified by a non-vanishing  $\langle P_{1,c} \rangle$ value. 
The smectic B polar (Sm$_{B,p}$) phase has the same organization of the standard Sm$_{B}$ phase with the additional feature of having azimuthal in-plane 
correlations, planes behaving independently from one another. 
This is dubbed as \textit{polar} 
to highlight this specific correlation, 
and is unscrewed ($\langle P_{1,c} \rangle=0$). 
Conversely, the screw-smectic B phase (Sm$_{B,S}^{*}$) is identified by a non 
vanishing value of the screw order parameter $\langle P_{1,c} \rangle$.

Table \ref{tab:1} summarizes the possible phases, as well as the corresponding set of order parameters that identify each of them.
\begin{table}[h]
\small
\begin{tabular*}{0.5\textwidth}{@{\extracolsep{\fill}}lll}
\hline
Phase & Code & Order parameter     \\
\hline
Nematic       & N                  & $\langle P_2 \rangle$ \\
Screw-nematic             & N$_{s}^{*}$        & $\langle P_2 \rangle$, $\langle P_{1,c}\rangle ^{(a)}$ \\
Smectic A                  & Sm$_{A}$           & $\langle \tau_1 \rangle$ \\
Screw-smectic A            & Sm$_{A,s}^{*}$     & $\langle \tau_1 \rangle$ , $\langle P_{1,c} \rangle ^{(a)}$   \\
Smectic B                  & Sm$_{B}$           & $\langle \tau_1 \rangle$, $\langle \psi_6 \rangle $ \\
Polar smectic B            & Sm$_{B,p}$         & $\langle \tau_1 \rangle$, $\langle \psi_6 \rangle$, $\langle P_{1,c} \rangle ^{(b)}$  \\   
Screw-smectic B            & Sm$_{B,s}^{*}$     & $\langle \tau_1 \rangle$, $\langle \psi_6 \rangle $, $\langle P_{1,c} \rangle^{(a)}$   \\
\hline 
\end{tabular*}
\caption{Summary of the different phases identified in the MC simulations of hard helices, along with the corresponding non-vanishing order parameters. $^{(a)}$: $\widehat{\mathbf{c}}$ rotates in helical way around $\widehat{\mathbf{n}}$; $^{(b)}$: $\widehat{\mathbf{c}}$ is defined for a each layer, without inter-layer correlation.}
\label{tab:1}
\end{table}

We can finally complement and enrich the information provided by the above order parameters with 
specific correlation functions that are designed to highlight the various trends observed.
The existence of  the nematic director $\widehat{\mathbf{n}}$
suggests the following decoupling of the
vector $\mathbf{R}_{ij} = \mathbf{R}_{j}-\mathbf{R}_{i}$ in components parallel $\mathbf{R}_{ij}^{\parallel}$ and perpendicular $\mathbf{R}_{ij}^{\perp}$ to $\widehat{\mathbf{n}}$
\begin{eqnarray}
\label{eq5}
\mathbf{R}_{ij}= \mathbf{R}_{ij}^{\parallel}+  \mathbf{R}_{ij}^{\perp}=\left(\mathbf{R}_{ij} \cdot \widehat{\mathbf{n}} \right) \widehat{\mathbf{n}}+
\left \vert \mathbf{R}_{ij} \times \widehat{\mathbf{n}} \right \vert \widehat{\mathbf{R}}_{ij}^{\perp}
\end{eqnarray}

Then, assuming $L_z$ as the box length parallel to $\widehat{\mathbf{n}}$ and $L_x,L_y$ as the other two, we can introduce the parallel  
\begin{eqnarray}
g_{\parallel} (R_{\parallel}) &=&
\frac{1} {N} \left \langle \frac{1}{\rho L_x L_y}  
\sum_{i=1}^N \sum_{j \neq i}^N  \delta (R_{\parallel} - 
\mathbf{R}_{ij} \cdot \hat{\mathbf{n}})  \right \rangle
\label{gpara}
\end{eqnarray}
and perpendicular
\begin{eqnarray}
g_{\perp} (R_{\perp})&=&
\frac{1} {2 \pi R_{\perp} N} \left \langle \frac{1}{\rho L_z}  
\sum_{i=1}^N \sum_{j \neq i}^N  \delta \left(R_{\perp} - 
\left \vert \mathbf{R}_{ij} \times \hat{\mathbf{n}} \right \vert \right)  \right \rangle \nonumber \\
\label{gperp}
\end{eqnarray}
positional correlation functions, that provide quantitative information on the spatial distribution of particles in the direction along $\widehat{\mathbf{n}}$ and in planes perpendicular to it.
Here $\rho=N/V$ is the number density, $V=L_x L_y L_z$ being the box volume. 
The onset of screw ordering is signalled by an oscillatory behavior of a special correlation function
\begin{eqnarray}
g_{1,\parallel}^{\widehat{\mathbf{w}}}(R_{\parallel}) &=& 
\left \langle \frac{\sum_{i=1}^N \sum_{j\neq i} ^ N \delta(R_{\parallel}-\mathbf{R}_{ij}\cdot \hat{\mathbf{n}}) (\widehat{\mathbf{w}}_i \cdot \widehat{\mathbf{w}}_j)} 
{\sum_{i=1}^N \sum_{j\neq i} ^ N \delta(R_{\|}-\mathbf{R}_{ij}\cdot \hat{\mathbf{n}})} \right  \rangle. \nonumber \\
\label{gw}
\end{eqnarray}
that measures the azimuthal correlation between the $\widehat{\mathbf{w}}$ axes of helices as a function of their distance along the  $\widehat{\mathbf{n}}$ director.

Typical behaviors of all the above quantities have been reported in Refs. \cite{Kolli14a,Kolli14b} and will not be reproduced here.
\section{Phase diagrams}
\label{sec:phase}
In the following we will present the calculated phase diagrams. 
In all cases, right-handed helical particles were taken. 
The same phase diagrams would be obtained for left-handed helices, 
with the only difference that the handedness of the chiral phases would be reversed.  
Scaled quantities are used throughout, with the bead diameter $D$ taken as the unit of length. 
Since the only interactions are hard-core repulsions, 
these systems are athermal and a single variable is sufficient to control the phase behaviour. 
Henceforth, the packing fraction $\eta$ will be used, which is defined as $\eta=\rho v_0$, 
where $v_0$ is the volume of a helix \cite{Frezza13}. 

\begin{figure}[htbp]
\begin{center}
\includegraphics[width=3.0in]{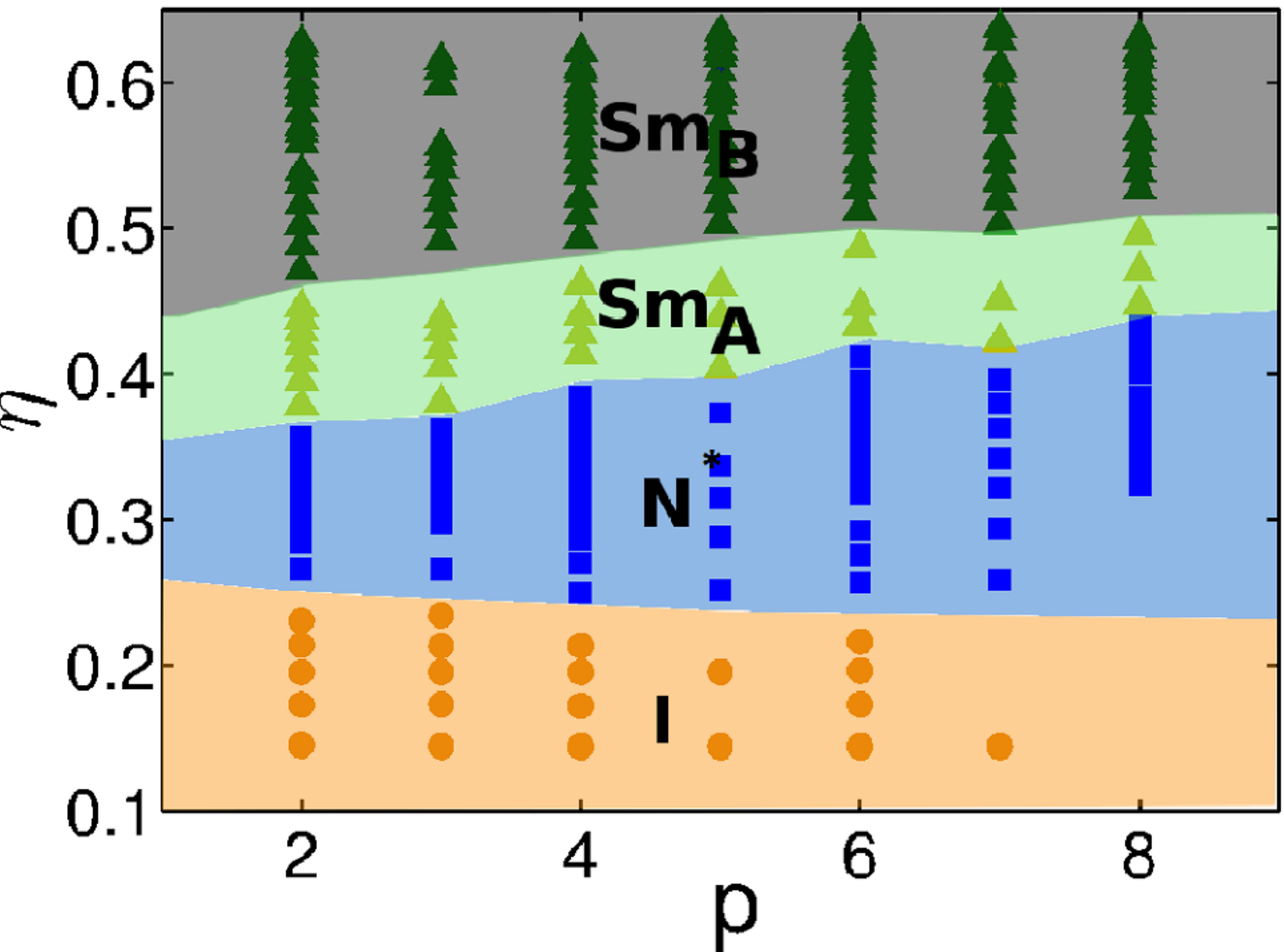}
\end{center}  
\caption{Phase diagram in the plane packing fraction $\eta$ versus particle pitch $p$ for helices with radius $r=0.1$.}
\label{fig:fig3}
\end{figure}

\begin{figure}[htbp]
\begin{center}
\includegraphics[width=3.0in]{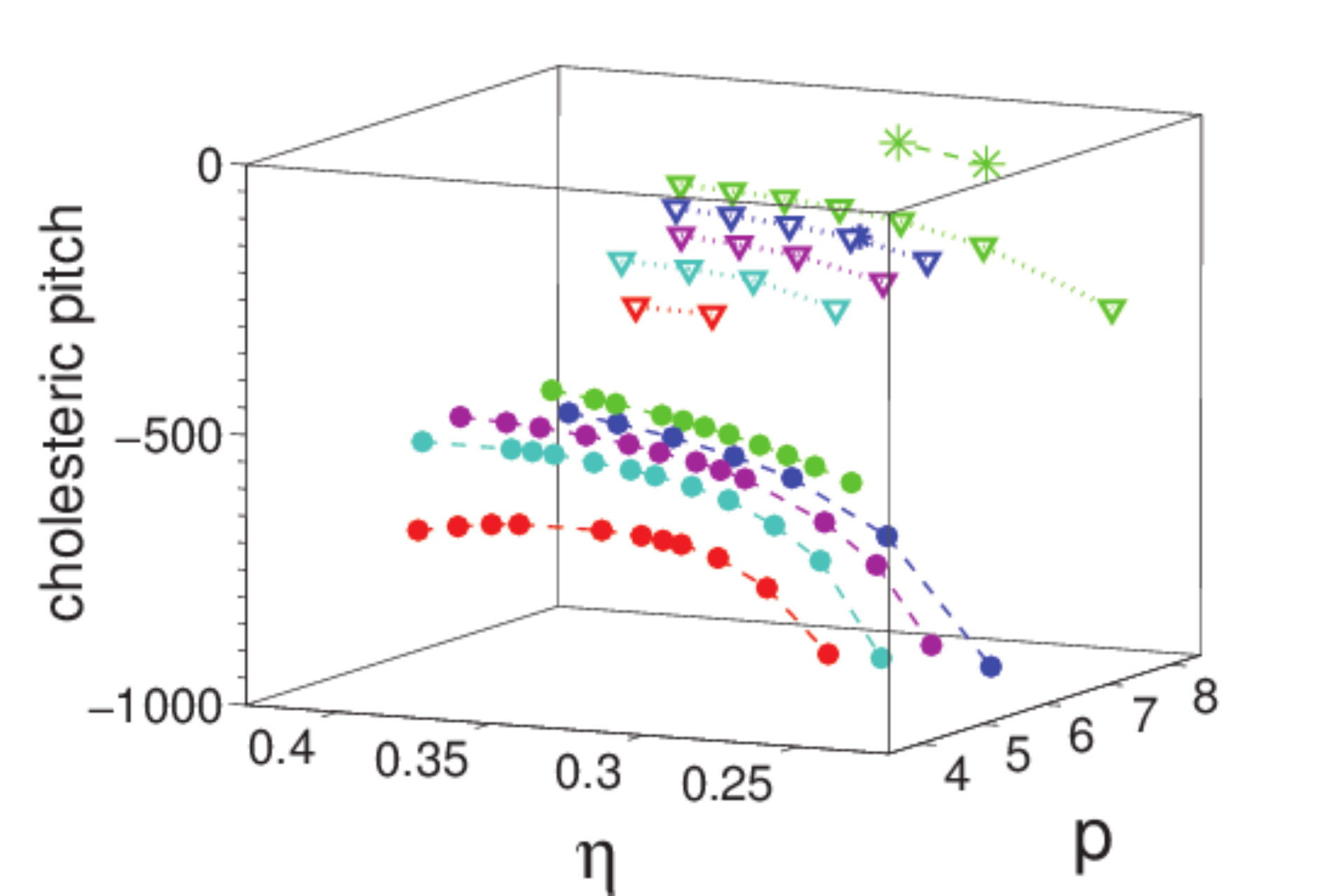}
\end{center}  
\caption{Cholesteric pitch $\cal P$, calculated for state points with $p>3$, in the N$^\ast$ phase of the phase diagrams in Fig. \ref{fig:fig3} ($r=0.1$, full circles), Fig. \ref{fig:fig6} ($r=0.2$, open triangles) and Fig. \ref{fig:fig9} ($r=0.4$ asteriscs). $\eta$ is the packing fraction and  $p$ is the helix pitch: $p=4$ (red), $p=5$ (cyan), $p=6$ (magenta), $p=7$ (blue), $p=8$ (green).}
\label{fig:fig4}
\end{figure}

\begin{figure}[htbp]
\begin{center}
\includegraphics[width=3.0in]{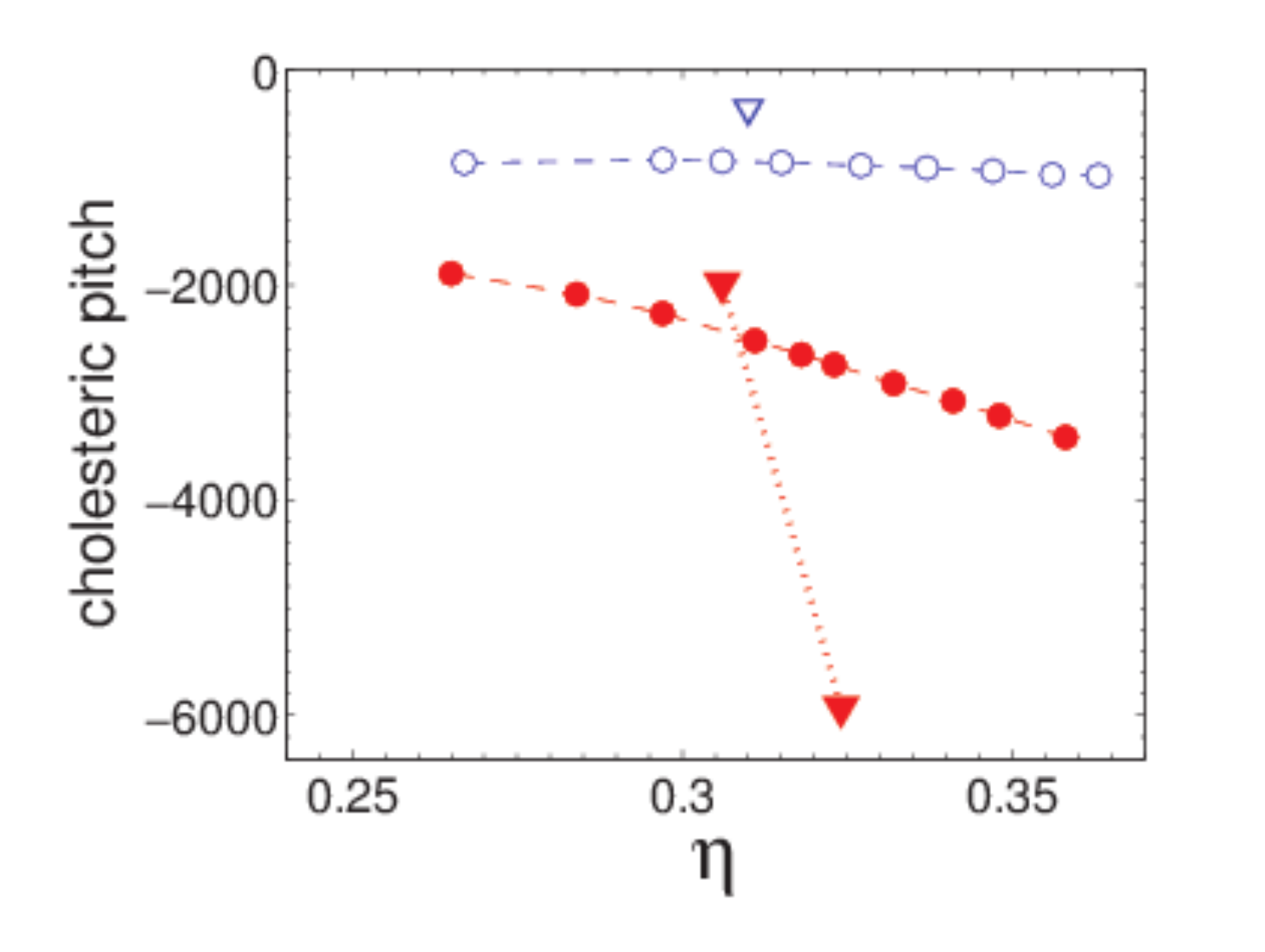}
\end{center}  
\caption{Cholesteric pitch $\cal P$ as a function of the packing fraction $\eta$, calculated for state points in the N$^\ast$ phase of the phase diagrams in Fig. \ref{fig:fig3} ($r=0.1$, circles) and Fig. \ref{fig:fig6} ($r=0.2$, triangles), for helices with pitch $p=2$ (full symbols) and $p=3$ (open symbols).}

\label{fig:fig5}
\end{figure}

We start our analysis with the simplest cases of very slender helices with $r=0.1$. 
Fig. \ref{fig:fig3} shows the phase diagram obtained for this case: 
irrespective of the particle pitch, on increasing density the sequence 
I $\rightarrow$  N $\rightarrow$ Sm$_A$ $\rightarrow$  Sm$_B$  is found, 
 as for hard spherocylinders having the same aspect ratio, $L/D  \sim 7.3-7.4$ \cite{Bolhuis97}. 
This example demonstrates that, 
if the helical character of particles is not sufficiently pronounced, 
which could be translated in terms of cavities that can be explored by neighboring particles,  
none of the helix distinct phases can be found. 
The only qualitative difference from 
the phase diagram of hard spherocylinders is 
in the low-density nematic phase, 
which  here is cholesteric rather than uniform nematic.
This is shown by  the values of $\cal P$ obtained from DFT calculations (Figs. \ref{fig:fig4} and  \ref{fig:fig5}).  
Cholesteric phases are not special to helices.
For all our right-handed helices with $r=0.1$ 
a left-handed N$^\ast$ organization is predicted (${\cal P} < 0$), 
with a pitch of the order of few hundreds for helices with $p>2$, 
which shows a dependence on density and tends to increase with decreasing $p$. 
A noticeable increase of the magnitude of $\cal P$, i.e. a decrease of the phase chirality, 
is found in the case of $p=2$.  
This can be explained by the fact that the in such helices the grooves are so tight that 
neighboring helices cannot easily lock into one  another. 
This is a general behavior, which was already pointed out \cite{Frezza14}: 
hard helices with very low pitch $p$ tend to form N$^\ast$ phases with large  periodicity $\cal P$, 
whose magnitude and sign are particularly sensitive to the packing fraction. 
 
The effects of the particle chirality on the phase diagram clearly appear when the helix radius increases to $r=0.2$, as shown in Fig. \ref{fig:fig4}. Considerable differences with the case $r=0.1$ are visible with new
phases almost all of which chiral. 
On increasing density, for all values of $p$, 
a transition from the isotropic to a cholesteric phase is found.
This,  according to DFT calculations, is  left-handed, 
with a density-dependent $\cal P$ in the range 200-400 for  $p>2$, 
and about an order of magnitude longer for $p=2$ (Figs. \ref{fig:fig4} and  \ref{fig:fig5}). 
These findings are not very different from those obtained for the case $r=0.1$.
However, the  N$^\ast$ phase has now a narrower stability range, especially for small $p$,
and is superseded, at higher density, by a   screw-nematic phase N$_{s}^{*}$  phase.
This is characterized by a helical rotation of the $C_2$ symmetry axis of helices ($\widehat{\mathbf{w}}$ vector in Figure \ref{fig:fig1}) around the the main director $\widehat{\mathbf{n}}$, which is the alignment direction of the helical axes $\widehat{\mathbf{u}}$.

Unlike the cholesteric, the N$^\ast_s$ phase has the same handedness (right) and the same pitch of the helical particles, 
essentially independent of the packing fraction.
The physical mechanism  of the transition to the N$^\ast_s$ phase  
was clearly identified in Refs. \cite{Kolli14a} and \cite{Kolli14b} as
 originating from a   coupling between the rotation of each helix around its own main axis 
(the unit vector  $\widehat{\mathbf{u}}$) and a translation along the nematic director $\widehat{\mathbf{n}}$. 
This coupling reduces the  
azimuthal freedom, and thus the rotational entropy, 
but such a loss is compensated by the increase of translational entropy 
allowed by the mutual sliding along $\widehat{\mathbf{n}}$ 
of neighboring parallel helices. 

\begin{figure}[htbp]
\begin{center}
\includegraphics[width=3.0in]{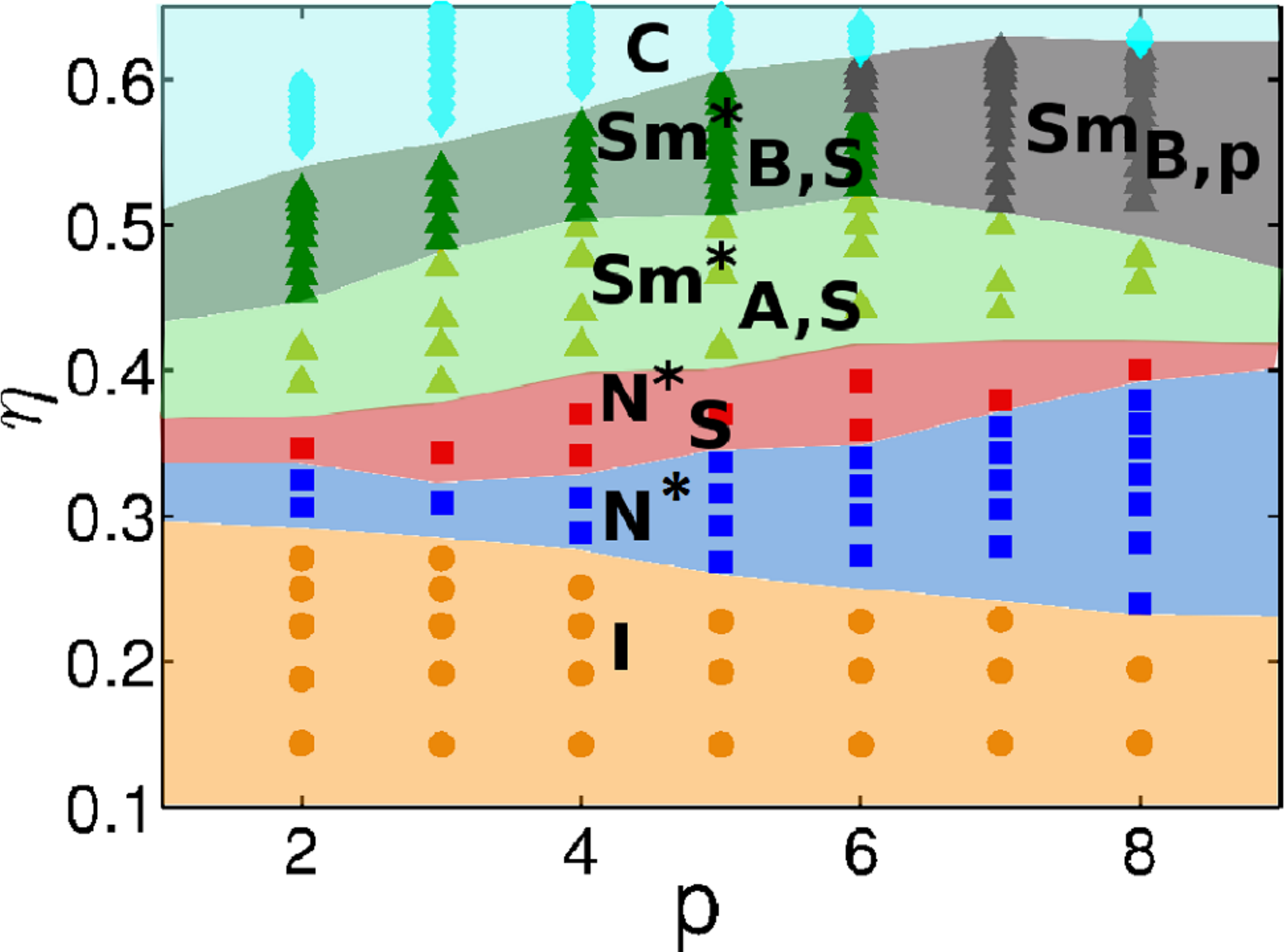}
\end{center}  
\caption{Phase diagram in the plane packing fraction $\eta$ versus particle pitch $p$ for helices with radius $r=0.2$.}
\label{fig:fig6}
\end{figure}

For helices with $r=0.2$ the screw-nematic N$^\ast_s$ phase is followed, at higher density, by a screw-smectic A phase 
(Sm$_{A,S}^{*}$)
which has screw order within layers. At even higher density two other smectic phases specific of helices appear, which have been denoted as screw-smectic B (Sm$_{B,S}^{*}$) and smectic B polar (Sm$_{B,p}$. The three smectic phases meet at a triple point located at $p \approx 6$ and $\eta \approx 0.5$.
\begin{figure}[htbp]
\begin{center}
\includegraphics[width=3.0in]{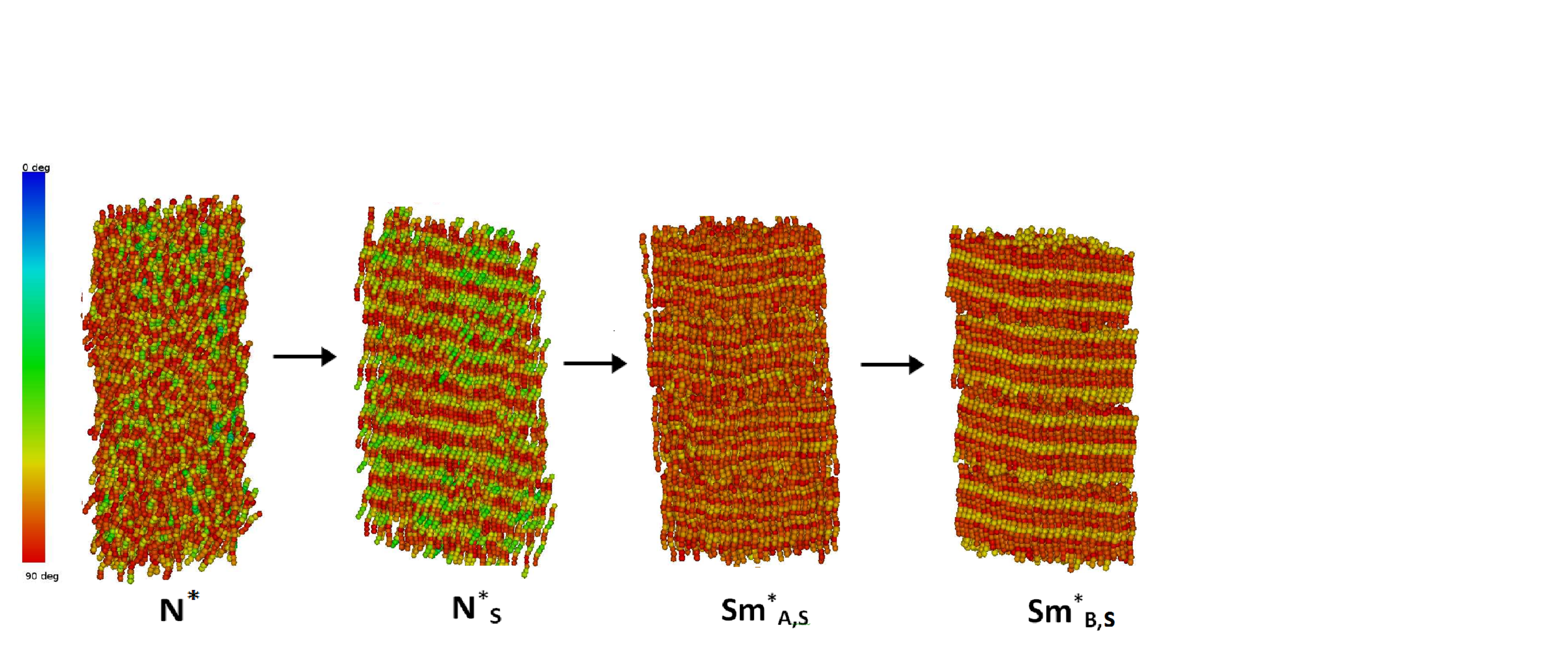}
\end{center}  
\caption{Snapshots from MC trajectories in the different liquid-crystalline phases for helices with  
$r=0.2$ and $p=4$. Color is coded according to the projection of the local tangent to helices onto a plane perpendicular to the director $\hat{\mathbf{n}}$ \cite{QMGA}. 
}
\label{fig:fig7}
\end{figure}

\begin{figure}[htbp]
\begin{center}
\includegraphics[width=3.0in]{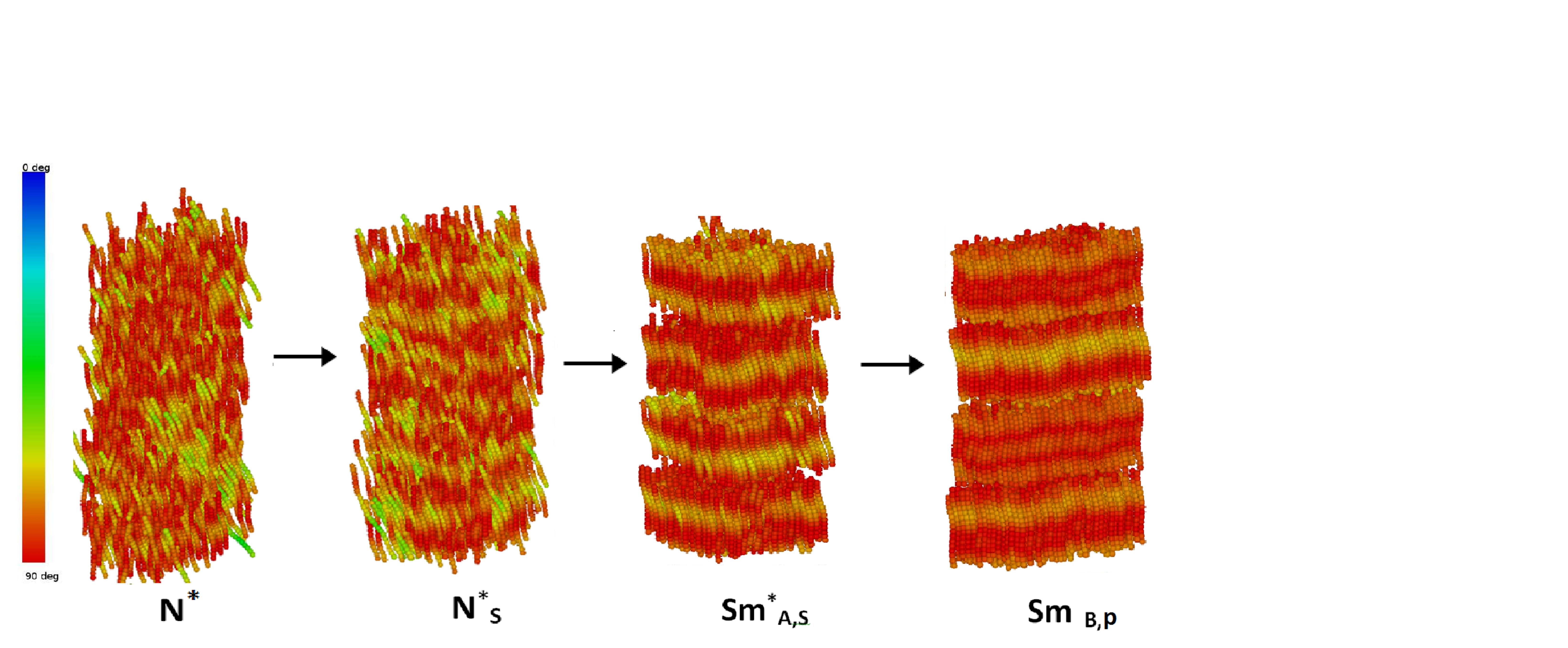}
\end{center}  
\caption{Snapshots from MC trajectories in the different liquid-crystalline phases for helices with  
$r=0.2$ and $p=8$. Color is coded according to the projection of the local tangent to helices onto a plane perpendicular to the director $\hat{\mathbf{n}}$ \cite{QMGA}. 
}
\label{fig:fig8}
\end{figure}

To give a visual insight of the  structure of the different phases, we show in Figs. \ref{fig:fig7} and \ref{fig:fig8} snapshots \cite{QMGA} taken from MC trajectories for helices with  $r=0.2$, $p=4$ 
and $p=8$, respectively. In the figures color is coded according to the projection of the local tangent to helices onto a plane perpendicular to the director $\hat{\mathbf{n}}$.  
So, the stripes that can be seen in the N$^\ast_s$ phase indicate the onset of azimuthal correlations between helices. The stripes persist in the Sm$^\ast_s$ phase, where there is the additional formation of layers. 
The regular spacing of the stripes indicates the presence of correlations between the layers: the helix $\hat{\mathbf{w}}$ axes rotate coherently in a screw-like  way across the layers. 
This visual indication can be translated in quantitative terms by computing the periodicity of $g_{1,\parallel}^{\widehat{\mathbf{w}}}(R_{\parallel})$ given in Eq.(\ref{gw}).

In the case of helices with $p=4$ the correlation between layers is maintained in the subsequent Sm$_{B,S}^{*}$ phase (Fig. \ref{fig:fig7}), which differs from the Sm$_{A,S}^{*}$ for the presence of short-range hexagonal ordering of the centers of mass of helices within the layers. Conversely, in the high density smectic phase of helices with  $p=8$ (Fig. \ref{fig:fig8}), the intra-layer correlation
is lost,  in favour of the possibility of layers
to freely slide and rotate  with respect to one another, in order to achieve the optimal packing, still maintaining the in-plane short range hexagonal ordering of the centers of mass, 
as well as the azimuthal coupling of neighbouring helices. The resulting smectic B phase Sm$_{B,p}$ is dubbed ``B polar'' 
to indicate the existence of  in-plane hexagonal
ordering (type B) and the in-plane azimuthal coupling (polar).

At higher densities, 
a more compact configuration generically labelled with ``C'' is found. 
Its characteristics are expected to be rather different from the crystal phases 
in the $r=0.1$ case (not shown in Fig.\ref{fig:fig3}), 
the latter being probably similar to those for hard spherocylinders. 
This is interesting in view of the possibility of forming crystals of specific shapes and chirality.

As the radius of the helix is further increased up to $r=0.4$, the screw-nematic phase N$_{S}^{*}$ widens
its stability range  at the expenses of the N$^\ast$ phase, as illustrated in Fig. \ref{fig:fig7}.
This is clearly due to the increased curliness of the $r=0.4$ case with respect to the $r=0.2$ counterpart (Fig.\ref{fig:fig1}), 
that favours the interlock intrusion of neighboring helices, 
thus promoting the azimuthal correlations responsible for the onset of the screw-nematic phase.
The disappearance of the colesteric phase at sufficiently small particle pitch, $p <3$, 
is a consequence of the small aspect ratio and reflects the lack of
a nematic phase for short hard spherocylinders \cite{Bolhuis97}. 
More interesting is the disappearance of the smectic B polar phase Sm$_{B,p}$ in favour of 
the screw-smectic B phase  Sm$_{B,S}^{*}$,
for any particle pitch above a certain packing fraction. 
This too can be rationalized in terms of the increased curliness of the helices that favours the screw order.

 For helices with  $p=7$ and $p=8$, a narrow cholesteric phase is predicted by DFT,  which is left-handed as for the helices $r=0.1$ and $r=0.2$, but has a much tighter pitch. This is one more consequence of the pronounced curliness of particles, 
which means high microscopic chirality. The predicted pitch $\cal P$, of the order of one hundred,
 is about 20 times the length scale of the particle chirality, thus much lower than typical ratios measured in cholesteric phases formed by molecular species, whether of low- or high-molar mass. 
Indeed, the high degree of microscopic chirality featured by helices with $r=0.4$ seems out of reach for molecular systems, also because of their intrinsic flexibility, 
which has the effect of reducing the net chirality \cite{review1}. 

Similarly to helices with $r=0.1$ and $r=0.2$ examined above, also in the case $r=0.4$ 
higher density configurations are characterized by compact "C" structures, 
at any particle pitch $p$, with notable exceptions of low particle pitches 
where plastic (rotor) phases are to be expected in view of the low aspect ratio, 
as in the case of hard spherocylinders\cite{Bolhuis97}.
\begin{figure}[htbp]
\begin{center}
\includegraphics[width=3.0in]{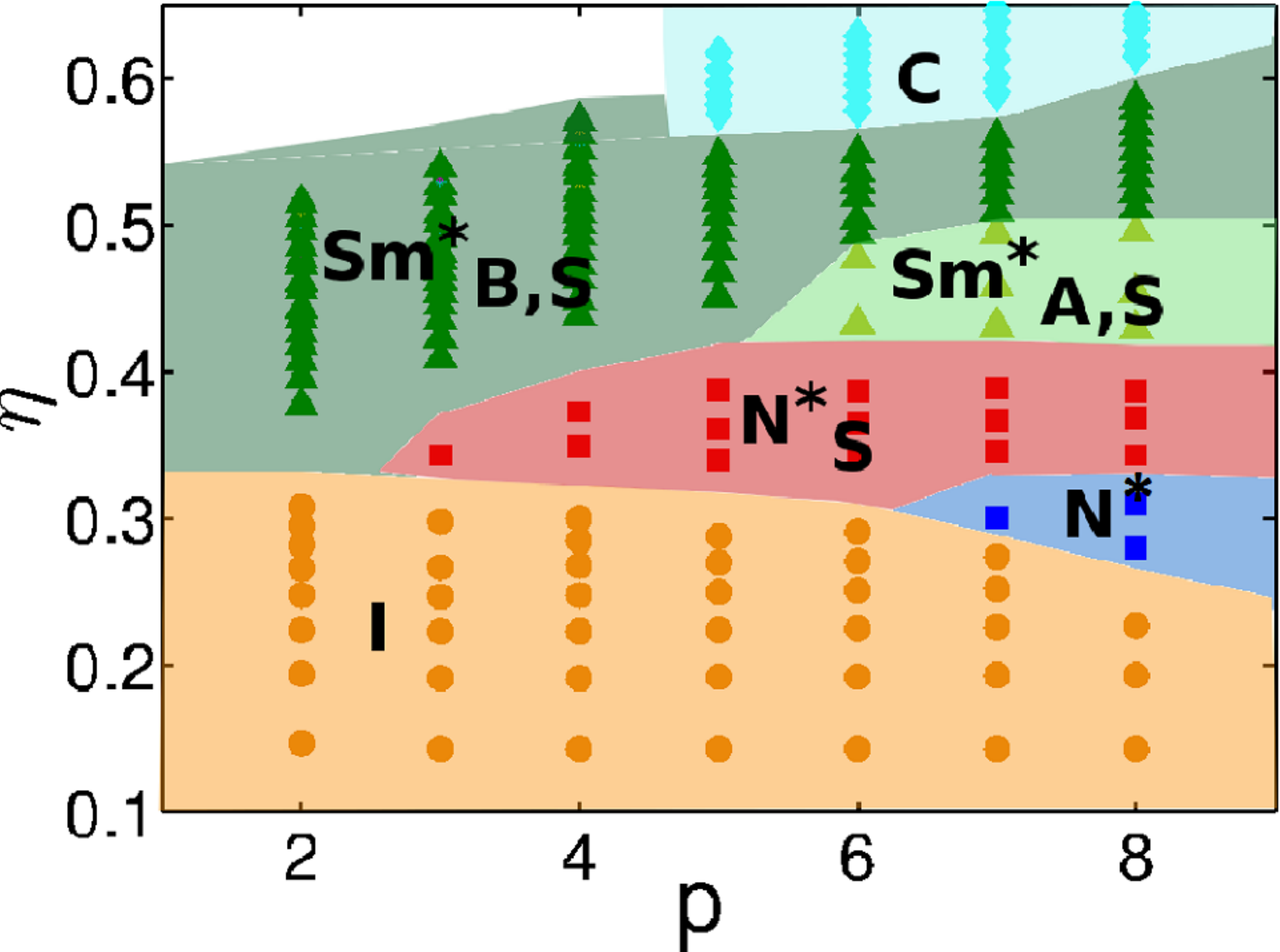}
\end{center}  
\caption{Phase diagram in the plane packing fraction $\eta$ versus particle pitch $p$ for  helices with radius $r=0.4$.}
\label{fig:fig9}
\end{figure}

\section{Discussion}
\label{sec:discussion}
In molecular systems, well-defined and rigid helices as our model particles seem challenging  
while the phase behaviour can be affected by the molecular flexibility \cite{lucaflexrods}.
If we would like to compare our predictions with experimental systems, 
then the best candidates appear to be colloidal suspensions of helical rigid particles.  
Indeed, screw-nematic order has been found in concentrated suspensions of helical flagella 
isolated from \textit{Salmonella typhimurium} \cite{Barry06}. 
While flagella with a rodlike shape exhibit a nematic phase, 
filaments with a pronounced helical character have been found to undergo a direct transition from the isotropic to the N$^\ast_s$ phase. The latter is the behaviour shown in Fig. \ref{fig:fig9} for most of the helices with $r=0.4$.
The experimental system did not exhibit any smectic phase, likely due to length polydispersity, 
as also suggested in the original work.
Length polydispersity is expected to promote further the stability of
the N$^\ast_s$ phase against the N$^\ast$ phase, 
as differently long helices with the same $p$ can all fit in the same N$^\ast_s$.
While potentially beneficial to the stability of the N$^{\ast}_s$ phase,
the length polydispersity is not essential to its formation,
as our results on monodisperse systems show. 
This is further supported by recent results of 
a parallel study on monodisperse hard helix systems
without employing PBC \cite{Cinacchi15} that confirm
the stabilisation of the N$^{\ast}_s$ phase over the N$^{\ast}$ phase at high density
and the absence of the latter in favour of the former for very curly helices, contrary to
what was alluded elsewhere \cite{Dussi15}.

The features of the N$^\ast_s$ phase found in our simulations and 
the fact that this is the same phase described in Ref. \cite{Barry06} were discussed in detail in \cite{Kolli14a,Kolli14b}.
We call this phase screw-nematic in line with Ref. \cite{Manna07}.
This was denoted as \emph{conical} in Ref. \cite{Barry06},
to account for the  conical arrangement of the local director,
identified with the local tangent to the helical particles,
as detected by polarization microscopy.
The experimental observations are fully consistent with our description of screw order in the N$_s^\ast$ phase,
with the helix axes ($\widehat{\mathbf{u}}$) preferentially aligned along the same direction
(the main director $\widehat{\mathbf{n}}$)
and the two-fold symmetry axes  of helices $\widehat{\mathbf{w}}$ spiralling around
$\widehat{\mathbf{n}}$.
We keep on with "screw-nematic" though, to distinguish this phase from
the \emph{conical} or \emph{heliconical} state induced 
in cholesterics by an applied electric or magnetic field parallel to
the twist axis.
In such a state, predicted almost 50 years ago \cite{MeyerdeGennes} and 
observed only recently \cite{Lavrentovich14}, 
the director  $\widehat{\mathbf{n}}$ performs a helical rotation around an axis at an oblique angle,
whence the name heliconical.
Unlike the N$_s^\ast$ phase, the heliconical state does not require the helical shape of constituents;
moreover, the periodicity of the heliconical state is similar to that of the underlying N$^\ast$ phase, 
i.e. orders of magnitude longer than the molecular size, and can be controlled by the applied field  \cite{Lavrentovich14}. 

Related but still different  is  the \emph{twist-bend} nematic (N$_{TB}$) phase, 
which was predicted for bent mesogens by theory \cite{Dozov} and simulations \cite{Memmer02}, 
and then experimentally observed only a few years ago \cite{Cestari,Chen,Borshch}.  
Like the N$_s^\ast$, the N$_{TB}$ phase has a small pitch helical modulation ($\cal P$ of the order of few molecular lengths), but it does not require helical, and not even chiral constituents. 
\section{Conclusions and perspectives}
\label{sec:conclusions}
The aim of the present study was to 
obtain a detailed account of the phase behaviour of hard helices  
in terms of their morphology. 
For the model systems shown in Fig. \ref{fig:fig1}, that resemble helical tubes of diameter $D$, 
the control parameters are the scaled radius $r/D$ and pitch $p/D$.
Using MC simulations, supplemented by DFT-Onsager second-virial calculations, we have obtained phase diagrams as a function of such parameters, thus providing a detailed picture of the structures that are formed as density increases. 
 
The phase diagrams are rich and exhibit a significant variety, even at a qualitative level, as 
the concavity of particles, which brings about the possibility of azimuthal correlations, becomes more pronounced. 
This is remarkable, especially in view of the absence of attractive  interactions. Since only hard-core repulsions have been assumed, all effects originate solely from changes in the morphology of the helices and the screw phases, not found in the corresponding hard rod counterpart, are entropically driven. 

For very slender helices, we find the sequence I $\rightarrow$ N$^\ast$ $\rightarrow$ Sm$_{A}$ $\rightarrow$ Sm$_{B}$  as 
in the phase diagram of hard spherocylinders \cite{Bolhuis97}, 
with the only difference that the uniform nematic  phase is replaced here by the chiral cholesteric phase, 
due to the chiral nature of particles. 
This shows that the helical character of particles has to be pronounced enough 
to promote the onset of phase organizations specific to helices, 
which are characterized by azimuthal correlations between particles. 
From the thermodynamic point of view, this means that only if the helical character is sufficiently pronounced, 
the entropy decrease due to the additional orientational order existing in such helix-specific phases can be 
compensated by the increase in translational entropy allowed by the roto-translational coupling. 
Based on these results, helical (bio)polymers, 
which in most cases  can be assimilated to very slender helices,
do not appear suitable to form helix-specific phases. 
This might explain why no experimental evidence of such phases have been given thus far 
for polymeric systems. 
However, the possibility that additional interactions (e.g. electrostatic) might enhance 
the helical character of real systems cannot be excluded.
Note also that the detection of screw distinct order in helical polymeric systems 
 is made difficult by the very small characteristic length scale 
that is of the order of the particle pitch.

Upon increasing the
radius of  helices from $r=0.1$ to $r=0.2$, a richer polymorphism is found and new phases appear. 
For $p \le 6$, we find a sequence I $\rightarrow $ N$^\ast$ $\rightarrow$ N$_{s}^{*}$ $\rightarrow$ Sm$_{A,s}^{*}$ $\rightarrow$ Sm$_{B,s}^{*}$, 
whereas for $6 < p \le 8$ the Sm$_{B,s}^{*}$ is replaced by a Sm$_{B,p}$ phase, indicating the existence of a triple point
between the  Sm$_{A,s}^{*}$,  Sm$_{B,s}^{*}$,  Sm$_{B,p}$ phases.
This trend is continued at $r=0.4$, with a direct sequence I $\rightarrow$ Sm$_{B,s}^{*}$ for $p \le 3$, I $\rightarrow$ N$_{s}^{*}$ $\rightarrow$ Sm$_{B,s}^{*}$ for $3< p \le 5$,
I $\rightarrow$ N$_{s}^{*}$ $\rightarrow$ Sm$_{A,S}^{*}$ $\rightarrow$ Sm$_{B,s}^{*}$ for $5< p \le 6$, and finally
I $\rightarrow$ N$^\ast$ $\rightarrow$ N$_{s}^{*}$ $\rightarrow$ Sm$_{A,s}^{*}$ $\rightarrow$Sm$_{B,s}^{*}$ above that value, with then several triple points in between. 
This very interesting phase behaviour requires highly helical particles and 
seems difficult to achieve in molecular systems. However colloidal suspensions of natural or artificial particles would be very good candidates to verify the theoretical predictions. We hope that our results can stimulate experimental work in this direction.
 
In all cases that we have examined here, right-handed helices were found to form a left-handed cholesteric phase and 
right-handed phases, either nematic or smectic, with screw ordering.  
In Ref. \cite{Dussi15}  examples of N$^\ast$ phase were reported, 
including a few cases examined here, 
exhibiting a change of   handedness from left to right with increasing density. 
However that study was limited to the cholesteric phase and 
did not account for the existence of the N$^\ast_s$ phase nor of any positionally ordered phase,
either by DFT \cite{notarellasme} or numerical simulation. 

Consideration of the whole phase diagram suggests that the change of N$^{\ast}$
handedness, although in principle possible, is fairly unlikely because of
the relatively small range of stability of the N$^\ast$ phase. For very
curly helices, on increasing density, the N$^\ast$ phase is superseded by
the N$^\ast_s$ phase before the inversion of cholesteric handedness has a
chance to take place. Such inversion is more likely to be observed for
large pitch cholesteric phases formed by slightly curly helices
\cite{Frezza14}, provided density is sufficiently low that positionally
ordered phases do not compete.

One interesting issue that originates from the present study would be
the analysis of the possible (if any) crystal phases of hard rigid helices,
that is a detailed characterization of the part of the phase diagram that
we have here generically labelled as ``C''.
In general, given a hard non-spherical particle model, 
it is not straightforward to predict
the structure of their crystal phases.
This holds particularly true if the particles are chiral.
In a very recent study \cite{Damasceno15}, this problem has been addressed by using 
an approach for shape design, where chiral entropic forces were tuned by
either shape unrouding or by an increase in depletant concentration. The use of helices  
where chiral entropic forces can be controlled in a very natural way by the particle morphology,
may represent a  more efficient way to control the
onset of a chiral crystal structure from the particle chirality.
We plan to address this issue in a future work,
possibly extending the methodology implemented in Ref. \cite{Kolli14b} and
combining it with insights stemming from   
Ref. \cite{Damasceno15}.
Equally of interest it would be to address the more general
question of the dense, ordered or disordered, packings of hard helices.
\begin{acknowledgments}
G.C. thanks the Government of Spain for
the award of a Ram\'{o}n y Cajal research fellowship and
the financial support under the grant FIS2013-47350-C5-1-R.
This work was also supported by MIUR PRIN-COFIN2010-2011 (contract 2010LKE4CC).
The use of the SCSCF multiprocessor cluster at 
the Universit\`{a} Ca' Foscari Venezia is greatfully acknowledged.
\end{acknowledgments}
%
\bibliographystyle{apsrev}

\begin{thebibliography}{99}
\bibitem{notarellasme}
Note that a study of positionally ordered liquid-crystal phases, like
smectic and columnar, can also be addressed by similar DFTs, e.g. in Ref. \cite{dftsme}.
\end{thebibliography}

\end{document}